\documentclass[10pt,a4paper,twoside,twocolumn,american,prd,nofootinbib,superscriptaddress]{revtex4-1}
\usepackage{lmodern}

\usepackage[T1]{fontenc}
\usepackage[utf8]{inputenc}
\usepackage{color}
\usepackage{babel}
\usepackage{booktabs}
\usepackage{units}
\usepackage{amsmath}
\usepackage{amssymb}
\usepackage{esint}
\usepackage[unicode=true,pdfusetitle,
 bookmarks=true,bookmarksnumbered=false,bookmarksopen=false,
 breaklinks=false,pdfborder={0 0 0},backref=false,colorlinks=true]
 {hyperref}
\hypersetup{
 citecolor=blue,filecolor=blue,linkcolor=blue,urlcolor=blue}

\makeatletter

\pdfpageheight\paperheight
\pdfpagewidth\paperwidth

\providecommand{\tabularnewline}{\\}

 \@ifundefined{textcolor}{}
 {%
   \definecolor{BLACK}{gray}{0}
   \definecolor{WHITE}{gray}{1}
   \definecolor{RED}{rgb}{1,0,0}
   \definecolor{GREEN}{rgb}{0,1,0}
   \definecolor{BLUE}{rgb}{0,0,1}
   \definecolor{CYAN}{cmyk}{1,0,0,0}
   \definecolor{MAGENTA}{cmyk}{0,1,0,0}
   \definecolor{YELLOW}{cmyk}{0,0,1,0}
 }

\renewcommand{\[}{\begin{equation}}
\renewcommand{\]}{\end{equation}}

\makeatother

\begin{document}

\title{Probing Dark Energy through Scale Dependence}

\author{Mariele Motta}

\affiliation{ITP, Ruprecht-Karls-Universität Heidelberg, Philosophenweg 16, 69120
Heidelberg, Germany}

\affiliation{Instituto de Física Gleb Wataghin -- UNICAMP, 13083-970 Campinas,
SP, Brazil}

\author{Ignacy Sawicki}

\affiliation{ITP, Ruprecht-Karls-Universität Heidelberg, Philosophenweg 16, 69120
Heidelberg, Germany}

\author{Ippocratis D. Saltas}

\affiliation{School of Physics \& Astronomy, University of Nottingham, Nottingham,
NG7 2RD, United Kingdom}

\author{Luca Amendola}

\affiliation{ITP, Ruprecht-Karls-Universität Heidelberg, Philosophenweg 16, 69120
Heidelberg, Germany}

\author{Martin Kunz}

\affiliation{Département de Physique Théorique and Center for Astroparticle Physics,
Université de Genève, Quai E.\ Ansermet 24, CH-1211 Genève 4, Switzerland}

\affiliation{African Institute for Mathematical Sciences, 6 Melrose Road, Muizenberg,
7945, South Africa}
\begin{abstract}
We consider the consequences of having no prior knowledge of the true
dark energy model for the interpretation of cosmological observations.
The magnitude of redshift-space distortions and weak-lensing shear
is determined by the metric on the geodesics of which galaxies and
light propagate. We show that, given precise enough observations,
we can use these data to completely reconstruct the metric on our
past lightcone and therefore to measure the scale- and time-dependence
of the anisotropic stress and the evolution of the gravitational potentials
in a model-independent manner. Since both dark matter and dark energy
affect the visible sector only through the gravitational field they
produce, they are inseparable without a model for dark energy: galaxy
bias cannot be measured and therefore the distribution of dark matter
determined; the peculiar velocity of dark matter can be identified
with that of the galaxies only when the equivalence principle holds.
Given these limitations, we show how one can nonetheless build tests
for classes of dark energy models which depend on making measurements
at multiple scales at a particular redshift. They are null tests on
the model-independent observables, do not require modeling evolution
in time and do not require any parametrization of the free functions
of these models---such as the sound speed. We show that one
in principle could rule out or constrain the whole class of the most-general
scalar-tensor theories even without assuming the quasi-static limit. 
\end{abstract}
\maketitle

\section{Introduction}

The introduction of dark energy relaxes many of the assumptions behind
$\Lambda$CDM cosmology. Not only can the background evolution be
altered, but in many models the dark energy is predicted to have significant
perturbations. As a result, the simple $\Lambda$CDM relations between
the dark matter perturbations and the gravitational fields sourced
by them are no longer valid. The detection of a form of dark energy
different from a cosmological constant hinges on the unambiguous determination
of how these relations are broken.

Various approaches to this goal have been taken. On the level of background
evolution, null tests for the compatibility of the cosmic expansion
history with $\Lambda$CDM were proposed in Refs \cite{Shafieloo:2009hi,Shafieloo:2011zv},
while a principal-component analysis of the equation of state constraints
was developed in Ref.~\cite{Huterer:2002hy}. On the level of linear
perturbations, parametrizations are usually used to give a model-independent
description of the growth of large-scale structure \cite{Caldwell:2005ai,Linder:2005in,Hu:2007pj,Amin:2007wi,Amendola:2007rr,Bean:2010zq},
although principal component analysis has also been performed \cite{Pogosian:2010tj}
(see \cite{Kunz:2012aw} for a short review). Alternatively one may
try to focus on understanding the limits that a general-relativity-like
theory must obey, but describing the remaining freedom with a parametrization
framework \cite{Bertschinger:2006aw,Battye:2012eu,Baker:2012zs} or
attempt to describe the properties of dark energy through an effective
framework \cite{Sawicki:2012re,Gubitosi:2012hu,Bloomfield:2012ff,Gleyzes:2013ooa}.
Such approaches, when contrasted with data, usually require parametrizations
in order to break down degeneracies, but which simultaneously introduce
parametrization-dependent biases. 

Our approach, introduced in Ref.~\cite{Amendola:2012ky}, has been
to take a step back and ask what the observables which we can in principle
extract from cosmological probes are when we do not assume a particular
model for dark energy. We have argued that as a result of the presence
of unknown dark energy perturbations, the only model-independent observables
are ratios of quantities typically considered to be unambiguous measurements.
This implies that such variables as the dark-matter growth rate $f$
or dark-matter perturbation amplitude $\sigma_{8}$ are not actually
observable in a model-independent way.

In this paper, we elucidate further this framework, showing that a
helpful way to understand our previous conclusions is to realize that
our cosmological probes (galaxy counts and weak lensing) rely on tracers
which propagate on geodesics and therefore can at best determine the
gravitational potentials which affect their motion. Indeed, we show
that given sufficient data one could in principle map out
the whole metric on our past lightcone. Therefore the slip parameter
describing anisotropic stress is a direct observable not requiring
any parametrization. Moreover, since it is a ratio of two
variables containing the same stochastic source, itself it is free
of cosmic variance. On the other hand, what we cannot do at all is
to ever probe the dark matter distribution using these means, since
the dark matter does not interact directly with our probes, only by
sourcing the gravitational potential. This means the effective Newton's
constant, which is important to quantify dark energy perturbations,
cannot be observed. Given this difficulty, we nonetheless describe
a strategy for constraining dark energy models, generalizing our work
away from the quasi-static limit considered previously.

We start in section \ref{sec:review} by defining the notation, and
reviewing the assumptions and the model-independent observables that
we have found before. We then demonstrate in section \ref{sec:obsprop}
that the metric and the gravitational slip can indeed be reconstructed
without further assumptions, but that the effective Newton's constant
is not fully determined. In section \ref{sec:testing} we illustrate
the practical consequences of this result by devising consistency
tests for different classes of models. For general dark energy models
in the quasi-static limit and dynamical Horndeski models with anisotropic
stress, we are able to formulate general constraints that relate different
scales at fixed redshifts and that can be tested by observations without
the need to parametrize any free functions. For models that have no
anisotropic stress, we propose a slightly more complicated procedure
using k-\emph{essence} as a worked example. We are again able to test
whether this model class is compatible with the data, and if it is,
a recipe allows one to reconstruct the model's free functions without
parametrization. We summarize the definitions of the variables used
and the status of their observability in table~\ref{tab:vars} located
on page \pageref{tab:vars}.

\section{Assumptions and Notation}

\label{sec:review}In Ref.~\cite{Amendola:2012ky}, we investigated
which properties of the constituents of the late universe are observable
without making prior assumptions as to the model of dark energy. We
started off with a minimal set of assumptions, which we require to
be able to interpret our observations at all: 
\begin{itemize}
\item The geometry of the universe is well described by scalar linear perturbations
in a Friedmann-Lemaître-Robertson-Walker metric with scale factor
$a(t)$, 
\[
\mathrm{d}s^{2}=-(1+2\Psi)\mathrm{d}t^{2}+a^{2}(t)(1+2\Phi)\mathrm{d}\boldsymbol{x}^{2}\,.
\]
We mostly neglect spatial curvature, but since it is observable, its
existence would not change the principal results of this paper. The
parameter $\Omega_{k0}$ would enter some of the equations, but given
how small it seems to be, it would not affect them in any significant
way. 
\item The matter content is pressureless. We do not differentiate between
dark matter and baryons. We neglect radiation. We assume that the
equivalence principle holds and therefore that both these components
follow geodesics of the same metric. This allows us to assume that
there is no velocity bias between dark matter and galaxies, although
we will comment on what breaking this assumption would mean.
\item The galaxy distribution is related to the matter distribution through
a potentially time- and scale-dependent deterministic linear bias,
$\delta_{\text{gal}}=b(k,a)\delta_{\text{m}}$. We make no assumption
about any particular form of the bias function. 
\end{itemize}
We have then assumed the idealized situation in which we can perform
current and future cosmological measurements to as good an accuracy
as is required for the range of scales and redshifts we are attempting
to characterize: supernova luminosities, the distribution of galaxies,
baryon acoustic oscillation measurements and weak lensing shear.

We have argued that without an \emph{a priori }assumption of a particular
model or parametrization for dark energy, there is no way of knowing
the evolution or even the initial conditions for the dark energy distribution.
Indeed, one of the aims of late-universe cosmology is exactly to measure
this, both on a background and perturbation level.

We shall use the standard notation to describe the evolution of the
cosmological background 
\[
E(z)^{2}=\Omega_{\text{m}0}(1+z)^{3}+\Omega_{k0}(1+z)^{2}+\Omega_{X}\,,
\]
with $E(z)\equiv H(z)/H_{0}$, the dimensionless Hubble parameter,
$\Omega_{\text{m0}}$ the contribution of matter to the total density,
$\Omega_{X}$ the contribution of dark energy and $\Omega_{k0}$ the
spatial curvature.

For the purpose of perturbations, we can describe the solution realized
by the Universe using a number of ratios of the gravitational potentials
in the metric---$\Phi$ and $\Psi$---and the dark matter configuration---the
density contrast $\delta_{\text{m}}$ and its peculiar velocity divergence
$\theta_{\text{m}}$. In all of the following, we denote with perturbation
variables (e.g.~$\Phi,\Psi,\delta_{\text{m}}$, etc.) the positive-definite
square root of their power spectrum, as is common in the literature.
We will always be working in Fourier space and primes will denote
a derivative with respect to $\ln a$.

In particular, we define ratios of the perturbation variables: the
slip parameter $\eta$, the effective dimensionless Newton's constant
$Y$, and the dark-matter growth rate $f$, which are all in principle
arbitrary functions of scale and redshift 
\begin{align}
\eta(k,z) & \equiv-\Phi/\Psi\,,\qquad Y(k,z)\equiv-\frac{2k^{2}\Psi}{3\Omega_{\text{m}}\delta_{\text{m}}}\,,\label{eq:PropFuncsDef}\\
f(k,z) & \equiv-\frac{\theta_{\text{m}}}{H\delta_{\text{m}}}\approx\frac{\delta_{\text{m}}'}{\delta_{\text{m}}}\nonumber 
\end{align}
where the scale $k$ has been expressed in units of the cosmological
horizon, $k\equiv k_{\text{com}}/aH$ and the prime denotes a derivative
with respect to $\ln a$. The perturbation variables are subject
to cosmic variance which will always put a limit on how precisely
their measurement reflects the actual variance. However, provided
that there is no additional source of stochastic perturbations in
the redshift range of interest, their \emph{ratios} for
a single mode $k$ are in fact free of cosmic variance (similar to
the argument in Ref.~\cite{Seljak:2008xr}). It is these ratios,
independent of the initial conditions, that we will use as variables
testing dark energy.

Our definition for $f$ is a little unusual, but it reduces to the
standard one sufficiently subhorizon, whenever the gravitational potentials
are slowly varying on the scales in question, since by the continuity
equation, 
\begin{equation}
\delta_{\text{m}}'+H^{-1}\theta_{\text{m}}=-3\Phi'\,.\label{eq:Continuity}
\end{equation}
We can also think of $Y$ as describing the size of the dark energy
perturbation relative to the one in matter, although adjusted for
anisotropic stress; at subhorizon scales and assuming no correlation
between matter and dark energy fluctuations, 
\begin{equation}
\eta Y\approx1+\frac{\Omega_{X}\delta_{X}}{\Omega_{\text{m}}\delta_{\text{m}}}\,.\label{eq:Yperts}
\end{equation}
One can form other ratios of all these variables, which are not independent.
In particular, a ratio useful for us will be 
\begin{equation}
\varpi\equiv\frac{f}{\Omega_{\text{m}0}Y}=\frac{3(1+z)^{3}\theta_{\text{m}}}{2HE^{2}k^{2}\Psi}\,.\label{eq:varpi}
\end{equation}
One should always bear in mind that these ratios in principle describe
the particular solution realized by our Universe rather than necessarily
representing fundamental properties of the dark energy. Both $Y$
and $\eta$ are unity on sub-horizon scales in $\Lambda$CDM, while
$f\approx\Omega_{\text{m}}^{0.55}$ in the concordance model. All
of these variables are scale independent subhorizon in $\Lambda$CDM.
This is not the case for most dark energy models.

It is important to stress that even measuring all these variables
does not give all the information on the configuration of dark energy.
In principle, there could be many internal degrees of freedom which
do not have an effect on the gravitational field produced (essentially
isocurvature modes). \\

In Ref.~\cite{Amendola:2012ky}, our focus was on determining the
properties of dark energy by attempting to extract from observations
its impact on dark matter. We first recovered the result of Ref.~\cite{Kunz:2007rk},
demonstrating that no combination of geometrical probes of background
evolution can ever split the contribution to total energy density
of the universe of matter and dark energy. These probes can at best
recover the Hubble parameter as a function of redshift up to an overall
normalization, $E(z)$, and independently measure any spatial curvature
$\Omega_{k0}$. Such properties as the dark energy equation-of-state
parameter, $w$, or $\Omega_{\text{m0}}$ are not observable, but
can only be obtained by making assumptions on the form of $w$ or
performing a particular parametrization. This degeneracy has been
dubbed the \emph{dark degeneracy}. Measurements of the local expansion
rate $H_{0}$ have been performed \cite{Riess:2011yx}. However, they
are sensitive to the local cosmic variance and therefore do not necessarily
reflect the averaged expansion rate today \cite{Marra:2013rba}.

We then extended this study to perturbations. We have demonstrated
that weak lensing and galaxy distribution observations can be reduced
to a measurement of three variables as function of scale and redshift
\begin{align}
A & =Gb\delta_{\text{m,0}}\,,\qquad R=Gf\delta_{\text{m,0}}\,,\label{eq:DirectObs}\\
L & =\Omega_{\text{m}0}GY(1+\eta)\delta_{\text{m,0}}\,,\nonumber 
\end{align}
where $G$ is the growth function normalized to one today, $\delta_{\text{m,0}}$
is the scale-dependent density contrast of dark matter. In Ref.~\cite{Amendola:2012ky}
we used the notation $\delta_{\text{m},0}=\sigma_{8}\delta_{\text{t,0}}$
to make explicit the fact that there is an unknown distribution of
dark energy perturbations present which affects gravitational measurements.
As we have previously argued, the amplitude of the density contrast
today depends on the whole history of the common evolution of the
unknown dark energy and dark matter, and the initial conditions in
both these components. It is thus impossible to measure $\delta_{\text{m,0}}$
without a knowledge of the bias $b$: the scale-dependence of $\delta_{\text{m,0}}$
is unknown without a knowledge of the dark energy model and there
are no measurements of the amplitude $\sigma_{8}$ that would be independent
of the DE model (see the discussion in section \ref{sec:Disc}). Thus
we have argued that the only model-independent observables are ratios
of the measurements \eqref{eq:DirectObs}: 
\begin{align}
P_{1} & =\beta=R/A=f/b\,,\label{eq:observables}\\
P_{2} & =E_{G}=L/R=\Omega_{\text{m0}}Y(1+\eta)/f\,,\nonumber \\
P_{3} & =R'/R=f+f'/f\,,\nonumber 
\end{align}
where for $P_{1}$ and $P_{2}$ we have provided also the notation
more frequently used by the community \cite{Zhang:2007nk}. It is
possible to measure each of these observables as a function of scale
and redshift. Any other such model-independent observable can be re-expressed
as a combination of the above and their derivatives. Indeed, a combination
that appears frequently is 
\[
\frac{L'}{L}=P_{3}+\frac{P_{2}'}{P_{2}}\,,
\]
which is not independent of the others. We will refer to the observables
$P_{i}$ as the \emph{primary model-independent observables}. The
observable $P_{1}$ contains the bias which cannot be measured in
a model-independent manner and therefore we will not use it any further
in our discussion (see the discussion of measurability of the bias
in section \ref{sec:Disc}). We should stress that these results,
for example, imply that the growth rate of dark matter $f$ cannot
be measured without picking a particular model or at least a parametrization
for dark energy.

We will now approach the question of observability slightly differently
than above, in a way that may be somewhat more physical. We will show
that what is actually observable in a model-independent manner are
the gravitational potentials, i.e.~the metric. What is not possible
to do using the observations we can perform is to split the sources
of the curvature into a dark matter and dark energy part.

\section{Directly Observable Properties \label{sec:obsprop}}

\label{sec:observables}We are going to begin this section by demonstrating
that using weak lensing (WL) and redshift-space distortions (RSD)
we can in principle reconstruct the metric in which the galaxies and
light propagate as a function of redshift and scale.

Weak lensing is a direct probe of the lensing potential 
\begin{equation}
k^{2}\Phi_{\text{lens}}=k^{2}(\Psi-\Phi)=-\frac{3(1+z)^{3}}{2E^{2}}L\,.\label{eq:Phi-Lens}
\end{equation}
Thus a tomographic weak lensing map should really be thought of as
a map of the lensing potential as a function of redshift and scale
\cite{Hu:1999ek}. It is not actually a direct measurement of the
underlying distribution of dark matter, which can only be extracted
by accounting for the unknown perturbations of dark energy also contributing
to the potential.

The above statement is quite natural and unsurprising. We will now
show that RSD can be thought of in the same way. Traditionally one
maps the velocities of the galaxy field onto the underlying dark-matter
distribution through the bias and relates the RSD to the dark matter
growth rate.%
\footnote{See e.g.~\cite{Kaiser:1987qv,Scoccimarro:1999ed}. This is how we
proceeded in Ref.~\cite{Amendola:2012ky}.%
} This is not necessary, however. Fundamentally, the Kaiser formula
\cite{Kaiser:1987qv} for the redshift-space galaxy density $\delta_{\text{gal}}^{z}$
is a statement about a correction to the real-space galaxy number
density $\delta_{\text{gal}}$ resulting from the peculiar velocities
of the galaxies $\theta_{\text{gal}}$ \cite{Scoccimarro:1999ed},
\begin{equation}
\delta_{\text{gal}}^{z}(k,z,\mu)=\delta_{\text{gal}}(k,z)-\mu^{2}\frac{\theta_{\text{gal}}(k,z)}{H}\,,\label{eq:Kaiser}
\end{equation}
with $\mu$ the direction cosine. Now, rather than relating this velocity
to dark matter evolution, we can make a choice that allows us to avoid
any assumption about the velocity bias: the galaxies are test particles
propagating on geodesics of the metric, whatever the actual source
of the curvature, 
\begin{equation}
\left(a^{2}\theta_{\text{gal}}\right)'=a^{2}Hk^{2}\Psi\,,\label{eq:rsd}
\end{equation}
with $k\equiv k_{\text{com}}/aH$. We can now integrate, neglecting
the integration constant, to discover that from the angular dependence
of the two-dimensional galaxy power spectrum we can extract, 
\[
A=\delta_{\text{gal}}\,,\quad R=-\frac{\theta_{\text{gal}}}{H}=-\left(a^{2}H\right)^{-1}\int a^{2}Hk^{2}\Psi\mathrm{d}\ln a\,.
\]
This approach allows us to see that linear RSD are really a measurement
of the gravitational potential $\Psi$ which accelerates the galaxies
\begin{equation}
-k^{2}\Psi=R'+R\left(2+\frac{E'}{E}\right)\,.\label{eq:PsiR}
\end{equation}

Since all the terms on the r.h.s.~of Eq.~\eqref{eq:PsiR} are observables,
we can use RSD to measure the gravitational potential as a function
of redshift and scale. Combining this with the information from the
lensing \eqref{eq:Phi-Lens}, allows us in principle to map out the
metric on our past light cone. This procedure will break down on scales
where the galaxies are no longer propagating on geodesics, i.e.~where
interactions become significant or non-linearities of the metric invalidate
the linear approximation. In principle, one should extend this simple
estimate for relativistic effects which will grow in importance closer
to the horizon, as in Refs \cite{Bernardeau:2009bm,Yoo:2009au,Challinor:2011bk,Bonvin:2011bg,Yoo:2012se}.

In models where both the baryons and the dark matter feel the same
forces (in particular, \emph{not }in e.g.~coupled quintessence \cite{Amendola:1999er}),
we can assume that there is no velocity bias between galaxies and
matter, thus $\theta_{\text{gal}}=\theta_{\text{m}}$. Through the
dark-matter continuity equation \eqref{eq:Continuity}, we can recover
the standard result \eqref{eq:DirectObs} that $R=\delta_{\text{m}}'$
for sufficiently slowly varying potentials, which was the basis of
our discussion in \cite{Amendola:2012ky}. When this assumption is
relaxed, which implies a violation of the equivalence principle, $\theta_{\text{m}}/H_{0}$
is no longer an observable because of the \emph{a priori }unknown
velocity bias. It will then no longer be true that RSD directly give
the derivative of the dark matter density contrast, $\delta_{\text{m}}'$.\\

We will now attempt to characterize the dark energy, which will require
us to return to the split between the dark matter and the dark energy.
Indeed, the total information that could potentially be extracted
from gravity-based experiments on the dark energy configuration realized
in our Universe is encoded in the ratios \eqref{eq:PropFuncsDef}.
We will now show that there is a fundamental difference in our ability
to extract the slip parameter $\eta$ from observations and the effective
Newton's constant, $Y$ or the growth rate $f$.

Firstly, since through Eqs \eqref{eq:Phi-Lens} and \eqref{eq:PsiR}
we can obtain the potentials, we can directly measure the slip parameter
by taking the appropriate ratio 
\begin{equation}
\eta=-\Phi/\Psi=\frac{3P_{2}(1+z)^{3}}{2E^{2}\left(P_{3}+2+\frac{E'}{E}\right)}-1\,.\label{eq:eta_obs}
\end{equation}
This is the same result as we obtained in Ref.~\cite{Amendola:2012ky}.
However, in the current approach, we have not had to make any assumptions
on the behavior of dark matter, or even its existence. Given sufficiently
precise measurements of the galaxy distribution and weak lensing covering
common redshifts and scales, the slip parameter can be measured redshift
by redshift, scale by scale, without having to depend on a better-
or worse-motivated parametrization and constraints that are obtained
through a total integrated signal.

Another result that will be useful in what follows is that, since
the potentials are observable, their evolution can also be measured.
We can define an observable akin to the growth rate for DM, the evolution
rate for the Newtonian potential, 
\begin{align}
\Gamma\equiv & \frac{\Psi'}{\Psi}=\frac{L'}{L}-\frac{\eta'}{1+\eta}-1\,,\label{eq:Potevol}\\
 & \frac{\Phi'}{\Phi}=\Gamma+\frac{\eta'}{\eta}=\frac{L'}{L}+\frac{\eta'}{\eta\left(1+\eta\right)}-1\,.\nonumber 
\end{align}
or equivalently, the evolution of $\Psi$ can obtained purely from
RSDs in terms of $P_{3}$ by differentiating Eq.~\eqref{eq:PsiR}.
Moreover, it is also possible to measure the variable $\varpi$ defined
in Eq.~\eqref{eq:varpi} 
\[
\varpi=\frac{1+\eta}{P_{2}}
\]
Again, the observability of $\varpi$ essentially depends on there
being no equivalence principle violation, since otherwise we do not
have access to $\theta_{\text{m}}$.\\

The measurement of anisotropic stress is a very important discriminator
between various classes of dark energy models. However, in order to
characterize the configuration of dark energy in our universe, we
also have to map the effective Newton's constant, $Y$, which describes
the relative size of the energy-density perturbations of dark energy
and matter, Eq.~\eqref{eq:Yperts}. Unfortunately, it is \emph{impossible
}to construct a combination of observables \eqref{eq:observables}
which would allow one to obtain a model-independent measurement of
$Y$. This is a result of the fact that there exists no cosmological
probe which would be able to determine the dark matter density $\delta_{\text{m}}$
without a prior assumption of a particular dark energy model. As we
have argued above, both lensing and RSDs probe the metric, with RSDs
at best providing a measurement of the derivative $\delta_{\text{m}}'$,
provided the potentials vary sufficiently slowly and the equivalence
principle be not violated. The best that we can do is to differentiate
the definition of $Y,$ Eq.~\eqref{eq:PropFuncsDef}, to obtain the
differential equation 
\begin{equation}
\frac{Y'}{Y}+\Omega_{\text{m}0}\varpi Y=1+\Gamma\,.\label{eq:Yobsevol}
\end{equation}
where we have dropped a term related to $\Phi'$ irrelevant subhorizon.
The coefficients in Eq.~\eqref{eq:Yobsevol} are observables (assuming
the equivalent principle), but we have only succeeded in obtaining
a first-order differential equation for $\Omega{}_{\text{m}0}Y$.
Given measurements of observables \eqref{eq:observables}, equation
\eqref{eq:Yobsevol} can be integrated. However, an initial condition
must be supplied, effectively the value of $\delta_{\text{m}}$ at
the initial time, which is not an observable.

The above limitation is again a result of the \emph{dark degeneracy},
here appearing on the level of perturbations. There is no gravity-based
experiment which could tell us in what part the local gravitational
potential is sourced by matter and in what part by dark energy. In
principle, one could hope to measure the mass density of dark matter
particles by using information from direct and indirect detection
experiments. However, since no experiment can currently determine
the physical density of dark matter on the scales of interest, the
split between dark matter and dark energy is fundamentally arbitrary.
We can always redefine the dark energy model to include a part of
the dark matter.%
\footnote{See for example Ref.~\cite{Lim:2010yk} where a model with no dark
matter, just dark energy, was proposed. The dark energy has sound
speed $c_{\text{s}}^{2}=0$ and can, for example, replicate \emph{exactly}
the phenomenology of the background and dark-matter linear perturbations
in $\Lambda$CDM.%
}

A part of the dark degeneracy is related to our lack of knowledge
of $\Omega_{\text{m}0}$ and implies that $\Omega_{\text{m}0}$ can
be absorbed into the definition of $Y$ through $Y\rightarrow\Omega_{\text{m}0}Y$.
However, this is not sufficient to fully specify the split between
dark matter and dark energy: we still have the freedom to supply an
arbitrary initial condition for \eqref{eq:Yobsevol}. For example,
one could imagine a situation where at some particular redshift there
are no dark energy perturbations but only those for matter, i.e.~$Y=1$.
Once this choice has been made, the observations in combination with
Eq.~\eqref{eq:Yobsevol} tell us how $Y$ evolves as a function of
redshift. However,\emph{ any }initial choice is as good as any other.
Eq.~\eqref{eq:Yobsevol} essentially evolves the values of $Y$ to
the next timeslice as a result of our assumption that the part of
the universe that we have designated to be dark matter is pressureless
and follows geodesics of a metric we have mapped.

As a result of this degeneracy, the best that we can do is to propose
a behavior for $Y$, resulting from some sort of model with some free
parameters, and then ask the question whether the observations are
consistent with the evolution mandated by Eq.~\eqref{eq:Yobsevol}.
Eq.~\eqref{eq:Yobsevol} should therefore be thought of as a \emph{consistency
relation} for $Y$ which can be used to rule out particular models
but not to measure this quantity. We will give an example of such
a procedure in section~\ref{sec:tests}.\\

The reason for this fundamental limitation is our lack of knowledge
of the density bias. The measurement of the amplitude of the galaxy
power spectrum only describes galaxy number counts; given a measurement
of the bias for some particular tracer, the dark matter density contrast
could also be reconstructed. Methods for the measurement of bias have
been proposed and used, but they assume various properties of $\Lambda$CDM
in a fundamental way. They cannot be used to constrain dark energy
models departing from the cosmological constant in general (see section
\ref{sec:Disc} for a discussion).\\

In light of the above results, we can reinterpret slightly the measurements
that we are performing in $\Lambda$CDM. The weak-lensing and galaxy-distribution
observations are in fact sensitive to the perturbed geometry, since
they involve tracers that propagate on geodesics. It is only because
the cosmological constant is not perturbed that these potentials can
be ascribed directly to dark matter and the standard language recovered.
It should be intuitively clear that without the knowledge of dark
energy perturbations we must have arrived at the model-independent
result we have described: only geometrical and not dynamical properties
can be measured by tracers propagating on geodesics.\\

The overall conclusion is that without particular assumptions on the
model of dark energy, cosmological observations allow at most for
the mapping out of the dimensionless Hubble parameter $E(z)$ as a
function of redshift and the two gravitational potentials as a function
of both redshift and scale. No information about the relative distribution
of dark energy and dark matter can be obtained without an assumption
of a particular model or at least some parametrization.

\section{Testing Scalar-Field Models \label{sec:testing}}

\label{sec:tests}In this section, we will discuss our ability to
use the observable properties described in section \ref{sec:observables}
to constrain or even exclude generalized scalar-tensor models of dark
energy.

\subsection{Quasi-Static Dark Energy}

\label{sec:QS}As we have already discussed in Ref.~\cite{Amendola:2012ky},
perturbations in scalar-tensor dark energy models frequently can be
approximated to evolve in the quasi-static approximation, where the
dark energy follows the dark matter perturbations and the time derivatives
are negligible (see also Ref.~\cite{Silvestri:2013ne}). For the
Horndeski Lagrangian \cite{Horndeski:1974,Deffayet:2011gz}, which
is the most general scalar-tensor theory involving no more than second
derivatives, this limit was derived in Ref.~\cite{DeFelice:2011hq}.
This derivation assumes that the only scales relevant to the problem
are the Jeans length (determined by the sound speed) and the Compton
wavelength, determined by the effective mass of the scalar perturbations.
Dispersion relations are in principle more complicated in these models,
but apart from the mass, the sound speed does provide the smallest
scale by the virtue of being defined in the $k\rightarrow\infty$
limit. Under the assumption that the speed of sound is close to that
of light and the scales larger than the Jeans length lie outside of
the scales probed by the observations, the effective Newton's constant
and the slip parameter \eqref{eq:PropFuncsDef} in these general scalar-tensor
models take the form 
\begin{equation}
Y=h_{1}\left(\frac{1+k^{2}h_{5}}{1+k^{2}h_{3}}\right)\,,\quad\eta=h_{2}\left(\frac{1+k^{2}h_{4}}{1+k^{2}h_{5}}\right)\,,\label{eq:Y+eta_QS}
\end{equation}
where the functions $h_{i}$ are purely functions of redshift determined
by the Lagrangian describing the dark energy model (see Ref.~\cite{Amendola:2012ky}
for the explicit form of $h_{1-5}$).

The quasi-static limit is really the requirement that the dark energy
perturbations follow the dark matter ones in a very constrained way.
Dark matter evolves on the geodesics of the combined gravitational
potential while the dark energy perturbation must follow the very
precise prescription defined by Eqs~\eqref{eq:Y+eta_QS}, without
any dynamics of its own.

One can use Eq.~\eqref{eq:eta_obs} to test the consistency of the
observed universe with the slip parameter $\eta$ given by the form
\eqref{eq:Y+eta_QS}. As we have described in Ref.~\cite{Amendola:2012ky},
given measurements of the observable \eqref{eq:eta_obs} at more than
three different scales per redshift, we can test whether the data
are consistent with a description for $\eta$ of the form \eqref{eq:Y+eta_QS}.
If not, then the anisotropic stress cannot be described as a manifestation
of a scalar-tensor theory in the quasi-static limit.

Since dark matter is supposed to move in a known way on geodesics
of the metric and we can map this metric out, the consistency of the
form of $Y$ given in the quasi-static limit can be tested, despite
the non-observability described in section \ref{sec:observables}.
Inserting the form \eqref{eq:Y+eta_QS} into the consistency relation
\eqref{eq:Yobsevol} allows one to ask, \emph{redshift by redshift},
whether the quasi-static form for $Y$ is consistent with the observational
data, i.e.~whether the simple scale dependence for $Y$ driven by
the coefficients $h_{i}$ is sufficient to explain the observations.
As a result of taking the time derivative in \eqref{eq:Yobsevol},
the number of free parameters increases to seven (including $\Omega_{\text{m}0}$),
thus a larger number of measurements at each redshift is required
to overconstrain the system, but it is in principle not any more difficult
than for the anisotropic stress.

The main takeaway from this discussion is that if the quasi-static
limit can be assumed, the dark energy configuration can be tested
by looking for a particular scale dependence of observables redshift
bin by redshift bin, rather than as a single integrated fit to the
observations with a particular parametrization chosen for the functions
$h_{i}$.

\subsection{Beyond the Quasi-Static Limit\label{sub:BeyQS}}

The quasi-static limit considered in section \ref{sec:QS} is somewhat
unsatisfactory, given that there are not necessarily \emph{a priori
}reasons why the DE should evolve so as to follow the DM perturbations
so closely. $Y$ specifies the size of the dark energy density perturbation
relative to the dark matter one. It should therefore be intuitively
clear that on any single spatial hypersurface, one can in principle
arrange the dark energy density in an arbitrary manner and therefore
there can be no restriction on the form of $Y$ if the dark energy
can be perturbed freely. By choosing the quasi-static limit, we give
up this freedom, since all the relative velocities between DM and
DE must vanish and the distribution of the DE is purely determined
by the density perturbation of dark matter.

Nonetheless, once the initial conditions are specified, $Y$ will
have to evolve in a manner consistent with the equation of motion
for the scalar or, equivalently, the conservation of the DE energy-momentum
tensor, whatever the model behind it. Since the metric is mapped out,
and the dark matter propagates on its geodesics, it is possible to
test whether the evolution of the dark energy part is consistent with
a particular class of models.

Another requirement arises in situations in which non-zero anisotropic
stress is detected. Its measurement through Eq.~\eqref{eq:eta_obs}
gives additional information on the distribution of the scalar-field
perturbations. This then has to be consistent with the distribution
implied by the effective Newton's constant $Y$ on that spatial slice.
Theories with anisotropic stress are in general more complicated than
those without and contain more parameters. However, in this way we
can have an additional handle with which to constrain them.

We will now discuss a couple of examples to illustrate how one might
go about constraining models using the model-independent variables.

\subsubsection{Consistency with k-essence}

\label{sub:kess-cons}Here, we will turn to the simplest class of
scalar field models, those described by the Lagrangian $\mathcal{L}_{\phi}=K(\phi,X)$
\cite{ArmendarizPicon:1999rj,ArmendarizPicon:2000dh,ArmendarizPicon:2000ah}.
The energy momentum tensor possessed by this kind of dark energy has
perfect-fluid form and at linear level all the properties are described
by the equation of state $w$ and the sound speed $c_{\text{s}}$
\cite{Garriga:1999vw}. This example will provide for sufficient complexity
in order to demonstrate the logic of our method and the fundamental
limitations of constraining the dark energy model space as a result
of the non-observability of $Y$, while having the advantage of being
familiar to a wide audience. Note that all uncoupled quintessence
models are contained within this class as are perfect-fluid models,
provided only scalar perturbations be considered.

We shall not give the full Einstein equations here, but it will suffice
to say that the combined k-\emph{essence} and dark matter energy-momentum
tensor (EMT) depends on the perturbation variables schematically as
\begin{align}
\delta T_{0}^{0} & \supset\delta\phi,\dot{\delta\phi}\,,\delta_{\text{m}}\,,\\
\delta T_{i}^{0} & \supset\delta\phi\,,\theta_{\text{m}}\,,\nonumber \\
\delta T_{i}^{i} & \supset\delta\phi,\dot{\delta\phi}\,,\nonumber \\
\delta T_{j}^{i}-\nicefrac{1}{3}\delta_{j}^{i}\delta T_{k}^{k} & =0\,,\nonumber 
\end{align}
with $\delta\phi$ the perturbation of the k-\emph{essence} scalar
and $\delta_{\text{m}},\theta_{\text{m}}$ describing the DM density
contrast and the divergence of the peculiar velocity. This model does
not allow for any anisotropic stress and therefore it could be immediately
excluded if, even at just one redshift and scale, $\eta\neq1$. We
will assume that no such detection was made and therefore we will
take $\Phi=-\Psi$. We can then use the combination of the Hamiltonian
and momentum constraints {[}i.e.~the $(00)$ and $(0i)$ Einstein
equations{]}, to eliminate the scalar-field perturbations $\delta\phi$
and $\dot{\delta\phi}$ in the $(ii)$ Einstein equation, obtaining
the exact equation for the evolution of the gravitational potential
in k-\emph{essence} models in the presence of dark matter 
\begin{align}
\Psi'' & +\left(4+\frac{E'}{E}+3c_{\text{a}}^{2}\right)\Psi'+\left(3+2\frac{E'}{E}+3c_{\text{a}}^{2}\right)\Psi+\label{eq:PsiEvol}\\
 & +c_{\text{s}}^{2}k^{2}\Psi=-\frac{3}{2}\Omega_{\text{m}}\left(c_{\text{s}}^{2}\delta_{\text{m}}+3(c_{\text{a}}^{2}-c_{\text{s}}^{2})k^{-2}H^{-1}\theta_{\text{m}}\right)\,.\nonumber 
\end{align}
where the adiabatic sound speed, 
\[
c_{\text{a}}^{2}\equiv\frac{\dot{p}_{X}}{\dot{\rho}_{X}}=-\frac{6\nicefrac{E'}{E}+2\left(\nicefrac{E'}{E}\right)^{2}+2\nicefrac{E''}{E}}{9\Omega_{\text{m}0}E^{-2}(1+z)^{3}+6\nicefrac{E'}{E}}\,,
\]
is fully determined by the observable expansion history up to $\Omega_{\text{m}0}$.
$\theta_{\text{m}}/H=R$ is of the order $k^{2}\Psi$ and is observable.
Equation \eqref{eq:PsiEvol} is quite standard, see the closely related
result in e.g.~\cite[Eq. 7.51]{Mukhanov}

Apart from the parameters of this model---the constant $\Omega_{\text{m}0}$
and the sound speed $c_{\text{s}}$, which can be a function of time---all
the quantities on the left-hand-side of Eq.~\eqref{eq:PsiEvol} are
observable. What is \emph{not }observable is $\delta_{\text{m}}$,
as we have explained in section \ref{sec:observables}. Since, at
any one redshift slice, the dark-matter configuration can in principle
be arbitrary, Eq.~\eqref{eq:PsiEvol} does not by itself provide
a constraint on the theory space. However, we can think of Eq.~\eqref{eq:PsiEvol}
as a measurement of $Y$ \emph{given the assumption }that the dark
energy model belongs to the k-\emph{essence} class: 
\begin{align}
c_{\text{s}}^{2}k^{2}\hat{Y}^{-1}= & \Gamma'+\left(\Gamma+3+\nicefrac{E'}{E}+3c_{\text{a}}^{2}\right)\left(1+\Gamma\right)+\label{eq:Y-kess}\\
 & +\nicefrac{E'}{E}+c_{\text{s}}^{2}k^{2}+3\Omega_{\text{m0}}(c_{\text{a}}^{2}-c_{\text{s}}^{2})\varpi\,,
\end{align}
where we have used Eq.~\eqref{eq:Potevol} to replace derivatives
of $\Psi$ with the observable $\Gamma$. So given observations of
$\Gamma$,$\varpi$ and $E$, $\hat{Y}\left[c_{\text{s}}^{2},\Omega_{\text{m}0}\right]$
is a functional on that data depending on the parameters $\Omega_{\text{m}0}$
and $c_{\text{s}}^{2}$ which outputs a function of scale and time.

However, we should stress that we can \emph{always} find \emph{some}
$\hat{Y}$ given \emph{any }data and given \emph{any }choice of parameters.
We have therefore thus far only obtained a \emph{model }for $Y$ in
the spirit of that provided by \eqref{eq:Y+eta_QS}. It is a little
more complicated, in that it depends not only on parameters equivalent
to the $h_{i}$'s, but also on observable data. If the observable
$\Gamma$ shows no scale dependence, in principle we have a very simple
model for $Y$, not dissimilar to the quasi-static one. If on the
other hand, the observations do show scale dependence, $\hat{Y}$
could be a very complicated function of scale.

We must now test whether this model for $Y$ is consistent with the
observations using the consistency relation \eqref{eq:Yobsevol},
just as we did in the quasi-static case. This means that k-\emph{essence}
can only be a good description of the dark energy if there is no anisotropic
stress observed and also 
\begin{equation}
\frac{\hat{Y}'}{\hat{Y}}+\frac{2\Omega_{\text{m}0}\hat{Y}}{P_{2}}=1+\Gamma\,,\label{eq:k-ess-Y}
\end{equation}
at every redshift and every scale. This consistency relation must
be valid given just one global parameter $\Omega_{\text{m}0}$ and
at each redshift the sound speed $c_{\text{s}}^{2}$ and its derivative.
Just as in the case of the quasi-static limit, at any one redshift,
given measurements at four scales it is in principle enough to exclude
such a model for $Y$ and therefore k-\emph{essence} as the mechanism
for dark energy, assuming the existence of a perfect data set, as
we already declared.%
\footnote{In fact, this would exclude any single perfect fluid as dark energy,
since k-\emph{essence} is equivalent to perfect-fluid hydrodynamics.
Multiple perfect fluids or fluids with internal degrees of freedom
have in general more complicated pressure perturbations, which would
naively appear as more complicated scale dependence (e.g.~\cite{Kunz:2006wc}). %
} Note that our approach is independent of the initial conditions.

Alternatively, we could have taken a time derivative of Eq.~\eqref{eq:PsiEvol}
and eliminated $\delta_{\text{m}}$ using the available equations
of motion. This gives a null test for k-\emph{essence}
purely in terms model-independent observables, but which involves
$\Psi'''$. This procedure is exactly equivalent to using the consistency
relation \eqref{eq:k-ess-Y}, which also involves an additional derivative.

If given the assumption of k-\emph{essence}, the consistency relation
is satisfied at all redshifts, then the above measurements can be
used to determine the value of the sound speed at each redshift observed.
This allows a non-parametric determination of this physical property
that in general is an arbitrary function of time.\\

It is worth mentioning the limit $c_{\text{s}}^{2}=0$ \cite{Creminelli:2009mu,Lim:2010yk},
which is not quasi-static on any scale. In such a case, the dust and
dark energy perturbations become indistinguishable and the entropy
perturbation source in Eq.~\eqref{eq:PsiEvol} disappears, yielding
\begin{equation}
\Psi''+\left(4+\frac{E'}{E}+3\tilde{c}_{\text{a}}^{2}\right)\Psi'+\left(3+2\frac{E'}{E}+3\tilde{c}_{\text{a}}^{2}\right)\Psi=0\,.\label{eq:DustPots}
\end{equation}
where $ $we have redefined the adiabatic sound speed to be that corresponding
to the total EMT, $\tilde{c}_{\text{a}}^{2}\equiv\dot{p}_{X}/(\dot{\rho}_{X}+\dot{\rho}_{\text{m}})$,
i.e.~it is now a function purely of the background geometry. In this
limit, there is no dependence on the unobservable $\delta_{\text{m}}$
and all scale dependence disappears. A measurement of $Y$ is not
possible in a fundamental sense, since there is no difference in the
properties of the DE and the DM perturbations. The dark degeneracy
is complete. The physics of the linear perturbations is effectively
that of a single generalized dust collapsing on a background with
some equation of state \cite{Lim:2010yk}. However, for such a model
to be a valid description of the observations, the data must satisfy
the constraint 
\[
\Gamma'+\left(\Gamma+3+2\frac{E'}{E}+3\tilde{c}_{\text{a}}^{2}\right)(\Gamma+1)+\frac{E'}{E}=0
\]
at every redshift and every scale with no free parameters. This constraint
reduces to exactly that of $\Lambda$CDM when the equation of state
for the dark energy becomes $-1$.\\

Models with kinetic gravity braiding (KGB) \cite{Deffayet:2010qz,Kimura:2010di}
are the most general class of scalar-tensor theories with a single
scalar which do not have a direct coupling to gravity and therefore
do not have anisotropic stress despite being imperfect fluids \cite{Pujolas:2011he}.
The KGB equivalent of Eq.~\eqref{eq:PsiEvol} would feature more
scales, however the prescription for constraining this class of models
would not differ from the k-\emph{essence} example presented here.
A null test for this class of theories would also involve
third derivatives of the potential.

\subsubsection{Consistency with general Horndeski theories featuring anisotropic
stress}

The most general second-order scalar-tensor theory is described by
the Horndeski Lagrangian \cite{Horndeski:1974,Deffayet:2011gz}. The
structure of the energy-momentum tensor for these theories is much
more complex than for a model such as k-\emph{essence}. In particular,
the Lagrangian\emph{ }features a non-minimal coupling to gravity.
This means that the EMTs for any such models will feature, in general,
anisotropic stress related to the perturbations of the scalar.

Horndeski models in the presence of dark matter have EMTs which have
the following dependence on the perturbed fields when only linear
perturbations are considered: 
\begin{align}
\delta T_{0}^{0} & \supset\dot{\delta\phi},\delta\phi\,,\delta_{\text{m}}\,,\\
\delta T_{i}^{0} & \supset\dot{\delta\phi},\delta\phi\,,\theta_{\text{m}}\,,\nonumber \\
\delta T_{j}^{i}-\nicefrac{1}{3}\delta_{j}^{i}\delta T_{k}^{k} & \supset\sigma\delta\phi\,,\nonumber \\
\delta T_{i}^{i} & \supset\ddot{\delta\phi},\dot{\delta\phi},\delta\phi\,,\nonumber 
\end{align}
in addition to dependence of all the components on the gravitational
potentials $\Phi$ and $\Psi$ and their time derivatives. Here $\sigma$
is fully determined as a function of the Horndeski free functions.%
\footnote{It is equivalent to the parameter $B_{7}$ in Ref.~\cite{DeFelice:2011hq}.
It is only non-zero when there is a direct coupling of the scalar
to gravity in the action.%
} As compared to k-\emph{essence}, we have the already mentioned anisotropic
stress, which is always proportional to $\delta\phi$. The perturbation
$\dot{\delta\phi}$ is present in the $(0i)$ components of the EMT,
while the pressure perturbation depends on $\ddot{\delta\phi}$. In
addition to the above, we have the equation of motion for the scalar
field, which is an equation for $\ddot{\delta\phi}$ in terms of all
the other variables. The full expressions can be found in Ref.~\cite{DeFelice:2011hq}.

We can clearly see that the only way to suppress the anisotropic stress
is to either make the coupling $\sigma$ very weak, i.e.~effectively
make the Horndeski terms non-minimally coupled to gravity irrelevant
for the dynamics of dark energy, or to suppress the scalar perturbations,
which is only possible if the scalar is very massive and not evolving,
i.e.~it just contributes vacuum energy \cite{Saltas:2010tt}.

Using the equation of motion for the scalar, we can eliminate $\ddot{\delta\phi}$
from the pressure equations. Then, using the Hamiltonian, momentum
and anisotropy constraints {[}i.e.~the $(00)$, $(0i)$ and $(ij)$
Einstein equations{]}, we can eliminate three further variables: the
scalar perturbations $\delta\phi,\dot{\delta\phi}$ and the dark-matter
density $\delta_{\text{m}}$. This is one extra variable as compared
to models without anisotropic stress, where the anisotropy constraint
equates the two potentials but is not dependent on the configuration
of the dark energy degree of freedom. The remaining form for the $(ii)$
Einstein equation is very simple, 
\begin{align}
\sigma\Big[\Phi'' & +\alpha_{1}\Phi'+\alpha_{2}\Psi'+\left(\alpha_{3}+\alpha_{4}k^{2}\right)\Phi\Big]+\label{eq:Horn-nov-1}\\
 & +\left(\alpha_{5}+\alpha_{6}k^{2}\right)\left(\Phi+\Psi\right)=\sigma\alpha_{7}\Omega_{\text{m}}k^{-2}\theta_{\text{m}}\,.\nonumber 
\end{align}
where the parameters $\alpha_{i}$ are functions of time alone and
are fully determined by the four free functions of the Horndeski Lagrangian.
Their exact form will not be useful here.

We have kept the dependence on $\sigma$ explicit to show that in
the limit of vanishing anisotropic stress, $\sigma\rightarrow0$,
Eq.~\eqref{eq:Horn-nov-1} is not an alternative dynamical equation
for the evolution of the potentials $\Phi$. To obtain the evolution
equation, one would have to eliminate $\Psi$ for $\Phi$ and would
obtain a very complicated equation with many scales embedded in its
coefficients. One may worry that Eq.~\eqref{eq:Horn-nov-1} provides
a different version of this evolution equation than that obtained
for theories with no anisotropic stress. This is not the case: switching
off the anisotropic stress reduces Eq.~\eqref{eq:Horn-nov-1} to
an equation which enforces $\Phi=-\Psi$, thus providing no additional
information.

Eq.~\eqref{eq:Horn-nov-1} is \emph{exact} at all scales where linear
perturbation theory is valid. In Eq.~\eqref{eq:Horn-nov-1}, all
the variables are model-independent observables. Using the extra information
from the anisotropic stress constraint, we have eliminated $\delta_{\text{m}}$,
which is not observable, and thus we can obtain a consistency relation
between pure observables required by all Horndeski models with anisotropic
stress,

\begin{align}
\eta\Gamma' & +\eta''+\Gamma\left(\eta\Gamma+2\eta'+\tilde{\alpha}_{1}\eta+\tilde{\alpha}_{2}\right)+\label{eq:HornConstr}\\
 & +\tilde{\alpha}_{1}\eta'+\tilde{\alpha}_{3}\eta+\tilde{\alpha}_{5}+k^{2}\left(\tilde{\alpha}_{4}\eta+\tilde{\alpha}_{6}\right)=\tilde{\alpha}_{7}\varpi\,.\nonumber 
\end{align}
where we have redefined the $\alpha_{i}$ to absorb factors of $\sigma$,
$E$ or $\Omega_{\text{m}}$, or to combine coefficients into single
variables.\\

A nice simple example of the relation \eqref{eq:Horn-nov-1} can be
obtained in the case of the $f(R)$ class of dark energy models, which
are a subclass of the Horndeski lagrangians and therefore feature
much less freedom \cite{DeFelice:2010aj}. Following the algorithm
described above, one obtains as the exact linear result 
\begin{align}
\Phi''-\Psi' & +\left(4+\frac{E'}{E}\right)\Phi'+\frac{1}{3}\left(m_{\text{C}}^{2}+2k^{2}\right)\Phi+\\
 & +\frac{1}{3}\left(m_{\text{C}}^{2}-6\left(2+\frac{E'}{E}\right)+k^{2}\right)\Psi=0\,,\nonumber 
\end{align}
where $m_{\text{C}}^{2}\equiv f_{,R}/2H^{2}f_{,RR}$ is the Compton
mass of the scalar degree of freedom in the units of Hubble. We must
have $m_{\text{C}}^{2}\gg1$ in order to satisfy Solar-System constraints
\cite{Hu:2007nk}. At small enough scales, $k\gg m_{\text{C}}$, we
can recover the standard quasi-static result in $f(R)$ that $2\Phi+\Psi=0$.
Again, we can express the above as a null test on observables that
needs to be satisfied at all scales where linear theory is valid 
\begin{align}
\eta\Gamma' & +\eta''+\Gamma\left(\eta\Gamma+2\eta'+5+\frac{E'}{E}\right)+\left(4+\frac{E'}{E}\right)\eta'+\nonumber \\
 & +\frac{m_{\text{C}}^{2}}{3}\left(\eta-1\right)+6\left(2+\frac{E'}{E}\right)+\frac{k^{2}}{3}\left(2\eta-1\right)=0\,.\label{eq:fR-constr}
\end{align}
This relation has only one free parameter, $m_{\text{C}}^{2}$, which
can be freely adjusted at each redshift to fit the data. If this is
not enough to satisfy the above test, then $f(R)$ is not a general
enough theory to account for the observations.\\

The relation \eqref{eq:HornConstr} {[}and \eqref{eq:fR-constr}{]}
is valid at all linear scales, without the need to involve the quasi-static
limit or choose initial conditions. It is enough to make a detection
of non-vanishing anisotropic stress at even just one scale and redshift
to then require that the data satisfy \eqref{eq:HornConstr}. At any
given redshift, the $\alpha_{i}$ coefficients are just numbers, independent
of scale. The observables, $\Gamma,\eta$ and $\varpi$ are in principle
complicated functions of scale and Eq.~\eqref{eq:HornConstr} is
a non-linear scale-dependent function on these data. Performing observations
for at least eight values of $k$ one can form an overconstrained
system and rule out the Horndeski model, or confirm it while measuring
its parameters $\alpha_{1-7}$, redshift by redshift without prior
parametrization \footnote{The parameters $\alpha_{1-7}$ are related to only four  
functions, in particular, those labelled as $w_i$ in Ref. \cite{Amendola:2012ky}. Therefore, they are are in principle, reducible to a smaller  
set of parameters.}.

\section{Discussion and Conclusions}

\label{sec:Disc}Dark matter is dark. This statement is obvious and
yet the discussion of cosmological observations is usually framed
as a discussion of the measurements of the dark-matter distribution.
This is of course completely natural in the case where there are no
dark energy perturbations, i.e.~within the framework of $\Lambda$CDM
cosmology. However, the moment that we relax this assumption, as we
must do when investigating dark energy models different from a cosmological
constant, this connection between measurements and the dark-matter
distribution is no longer simple, and can to all intents and purposes
disappear.

In this paper, we have stressed that the probes we hope to use to
constrain general dark energy models depend on tracers which propagate
on geodesics: light for weak lensing and galaxies for RSD. This allows
us to map out the metric on our past lightcone (the potentials $\Phi$
and $\Psi$) through which these tracers fall freely. These are the
real physical observables of the combination of the cosmological probes.
The relation of these potentials to the dark-matter distribution is
then a model-dependent statement, which just happens to be an identity
for $\Lambda$CDM.

Given this limitation, we should stress that the cosmological probes
still can provide an enormous amount of information. Since we can
reconstruct the potentials, the slip parameter $\eta$ of the matter/light
Jordan-frame metric is a model-independent observable, given overlapping
WL and RSD measurements. We do not need to parametrize it and we can
answer the question of whether anisotropic stress is at all necessary,
directly from observations, without any further modeling. This test
has the power to immediately eliminate very large classes of models,
whatever its result. If $\eta\neq1$ at even one redshift and scale,
then models without non-minimal coupling to gravity in the baryons'
Jordan frame can be thrown away. If $\eta=1$ everywhere, then such
couplings of dark energy to gravity could in principle still be there,
but they are so small that they cannot at all influence the dynamics
of the universe and therefore can be neglected. We have summarized
the most important dynamical variables and their observability status
in Table~\ref{tab:vars}.

We have also shown that the evolution rate of the \emph{potentials}
can be mapped out. On the other hand, the fact that we cannot observe
the dark matter density amplitude means that the effective Newton's
constant $Y$ is not an observable in a model-independent setting.
Given these model-independent observables, we should refocus the predictions
of perturbation theory in dark energy models to their impact on these
gravitational potentials rather than on the dark matter, contrary
to the dominant approach in the discussion today. 

Nonetheless, we have demonstrated in section \ref{sec:tests} how
to construct tests of very general classes of models of dark energy,
even away from the frequently employed quasi-static limit. The distinct
advantage of this approach is that rather than parametrizing the free
functions of a class of models first, then evolving the predictions
of these models using modified codes and simultaneously fitting to
data, our method is capable of testing/constraining classes of models
without any prior parametrization and without any assumptions on the
initial conditions. These tests involve performing measurements of
the observables at multiple scales, redshift by redshift (with the
number of data points required determined purely by the number of
free functions of time in the particular class of models). In order
to carry out this test it is not necessary to modify codes for each
class of models, but just to apply appropriate transformations to
the observed data. In other words, our approach calls for exploring
the space domain rather than the time domain of dark energy.

We have not addressed here the issue of whether such a method
is indeed practical, delaying detailed forecasts to a separate follow-up
work. A measurement of the slip parameter using Eq.~\eqref{eq:eta_obs}
may not ever provide a better test for particular models of
DE than the standard method. However, this test uniquely is model
independent and it is not immediately obvious that the use of derivatives
of data decreases information when we benefit by being able to construct
a null test which must be satisfied at every scale and every redshift
if $\Lambda$CDM is to not be excluded, especially if cosmic variance
can be combatted.

The tests constructed beyond the quasi-static limit in section~\ref{sub:BeyQS}
do appear significantly less realistic. However, the standard
methodology of parameterizing $\eta$ and $Y$ does not even address
what might occur in situations where the quasi-static limit does not
apply. Moreover, given the lack of well-motivated choices of potentials
and initial conditions for DE models, eliminating a \emph{class}
of models, e.g.~quintessence, as the mechanism for DE would significantly
widen constraints since multiple potentials would have to be tested.
It is thus unclear whether any other, more standard method would perform
any better.

\begin{table*}[th]
\begin{tabular}{lllllrrrl}
\toprule 
\textsf{\textbf{\small{Class}}}{\small{ }} &  & \textsf{\textbf{\small{Variable}}}{\small{ }} &  & \textsf{\textbf{\small{Key Relation}}}{\small{ }} & \textsf{\textbf{\small{Obs.?}}}{\small{ }} & \textsf{\textbf{\small{Eq.}}}{\small{ }} &  & \textsf{\textbf{\small{Comment}}}\tabularnewline
\midrule
\midrule 
\emph{\small{Measurements}}{\small{ }} &  & {\small{$A$}}  &  & $\equiv\delta_{\text{gal}}=b\delta_{\text{m}}$  & $\checkmark$  & \eqref{eq:DirectObs}  &  & Bias not measurable absent DE model\tabularnewline
\cmidrule{3-9} 
 &  & {\small{$R$}}  &  & $\equiv-\theta_{\text{gal}}/H$  & $\checkmark$  & \eqref{eq:DirectObs}  &  & RSD measure galaxy velocity\tabularnewline
\cmidrule{3-9} 
 &  & {\small{$L$}}  &  & $\equiv\frac{2k^{2}E^{2}}{3(1+z)^{3}}(\Phi-\Psi)$  & $\checkmark$  & \eqref{eq:DirectObs}  &  & WL shear probes lensing potential\tabularnewline
\cmidrule{3-9} 
 &  & $E$  &  & $\equiv H/H_{0}$  & $\checkmark$  &  &  & $H_{0}$, $\Omega_{\text{m0}}$ not observable without DE model%
\footnote{See Ref.~\cite{Amendola:2012ky} for a detailed discussion of the
observability of the background.%
}\tabularnewline
\midrule 
\emph{Primary}  &  & $P_{1}=\beta$  &  & $\equiv R/A=f/b$  & $\checkmark$  & \eqref{eq:observables}  &  & \tabularnewline
\cmidrule{3-9} 
\emph{observables}  &  & $P_{2}=E_{G}$  &  & $\equiv L/R=(1+\eta)/\varpi$  & $\checkmark$  & \eqref{eq:observables}  &  & \tabularnewline
\cmidrule{3-9} 
 &  & $P_{3}$  &  & $\equiv R'/R=f+f'/f$  & $\checkmark$  & \eqref{eq:observables}  &  & Only this function of $f$ is observable\tabularnewline
\midrule 
\emph{\small{Physical }}{\small{ }} &  & {\small{$\Psi$}}  &  & $R'+R\left(2+\nicefrac{E'}{E}\right)$  & $\checkmark$  & \eqref{eq:PsiR}  &  & Extract from RSD\tabularnewline
\cmidrule{3-9} 
\emph{\small{variables}}{\small{ }} &  & {\small{$\Phi$}}  &  & $\frac{3(1+z)^{3}}{2k^{2}E^{2}}L+\Psi$  & $\checkmark$  &  &  & Extract from WL tomography\tabularnewline
\cmidrule{3-9} 
 &  & {\small{$\delta_{\text{m}}$}}  &  & $\delta_{\text{m}}=\delta_{\text{gal}}/b$  &  &  &  & Unknown without knowing bias\tabularnewline
\cmidrule{3-9} 
 &  & {\small{$\theta_{\text{m}}/H_{0}$}}  &  & $\theta_{\text{m}}=\theta_{\text{gal}}$  & $\checkmark$  &  &  & Observable only given equivalence principle\tabularnewline
\midrule 
\emph{\small{DE }}{\small{ }} &  & {\small{$\eta$}}  &  & $\equiv-\Phi/\Psi$  & $\checkmark$  & \eqref{eq:PropFuncsDef}  &  & $\eta\neq1\Rightarrow$ coupling DE/gravity\tabularnewline
\cmidrule{3-9} 
\emph{\small{Configuration}}{\small{ }} &  & {\small{$\Gamma$}}  &  & $\equiv\Psi'/\Psi$  & $\checkmark$  & \eqref{eq:Potevol}  &  & Observable prediction of DE models\tabularnewline
\cmidrule{3-9} 
\emph{\small{Variables}}{\small{ }} &  & $Y$  &  & $\equiv-2k^{2}\Psi/3\Omega_{\text{m}}\delta_{\text{m}}$  &  & \eqref{eq:PropFuncsDef}  &  & $\delta_{\text{m}}$ unknown; must satisfy relation \eqref{eq:Yobsevol}\tabularnewline
\cmidrule{3-9} 
 &  & $f$  &  & $\equiv-\theta_{\text{m}}/H\delta_{\text{m}}\approx\delta_{\text{m}}'/\delta_{\text{m}}$  &  & \eqref{eq:PropFuncsDef}  &  & $\delta_{\text{m}}$ unknown\tabularnewline
\cmidrule{3-9} 
 &  & $\varpi$  &  & $\equiv f/\Omega_{\text{m}0}Y$  & $\checkmark$  & \eqref{eq:varpi}  &  & Relates DM velocity to potential\tabularnewline
\bottomrule
\end{tabular}\caption{\label{tab:vars}Summary of variables used in this paper. Tshose variables
marked with a checkmark as observable can be measured as a function
of redshift and scale without an assumption of a particular DE model.
Measurement of $\delta_{\text{m}}$ requires the knowledge of bias,
which cannot be measured or modeled without the knowledge of the DE
model. The variables such as $Y$ and $f$ typically used to describe
the DE configuration are therefore not observable, but predictions
can be reformulated to maximally exploit those variables which are
observable (see section \ref{sub:BeyQS}).}
\end{table*}

We have mostly assumed that there is no velocity bias between dark
matter and the galaxies. Thus a measurement of galaxy peculiar velocities
$\theta_{\text{gal}}$ via RSD is automatically a measurement of the
dark-matter peculiar velocity $\theta_{\text{m}}$. Statistical origins
of the velocity bias notwithstanding \cite{Desjacques:2009kt}, there
could in principle be a non-vanishing bias between these two velocities.
Indeed, this happens in any model in which there is a fifth force
acting on dark matter, which does not couple to baryons or light,
i.e.~models with a violation of the equivalence principle such as
coupled quintessence \cite{Amendola:1999er}. This fifth force is
a new source of acceleration for dark-matter particles, causing their
peculiar velocities to deviate from that of the galaxies. The effect
of this is to introduce a (scale and time-dependent!) velocity bias.
Our measurements of WL and RSD still map out the same metric potentials
for our tracers, but these are not the complete potentials that the
dark matter feels. Allowing for such deviations, we completely lose
the connection between the dark matter growth rate and RSD of the
galaxy power spectrum.

Summarizing the above, we can say that RSD and WL map out the Jordan-frame
metric for galaxies and light, but provide no direct information on
the Jordan frame of the underlying dark matter. In principle, one
could use different tracers with different baryon fractions (cluster
and galaxy power spectra) to map out the Jordan frames for these two
classes of objects and obtain some information on the differences
in the gravitational potentials they experience (in the spirit of
e.g.~\cite{Hui:2012jb}). Interestingly, galaxies are not just baryons,
but are dynamically coupled to their dark-matter halos. This means
that the Jordan-frame metric for the galaxies is not necessarily the
same as the one for baryons, since the galaxies will at least partly
feel the fifth force acting on dark-matter.\\

Another interesting implication of our approach is that potential
non-linearities of dark energy, which can appear on very different
scales to those in dark matter as a result of the generic existence
of screening mechanisms, are not deadly to the measurements of the
linear potentials $\Phi$ and $\Psi$. For tracers to move on geodesics
determined by these potentials, their gradients must remain small,
i.e.~the \emph{total }EMT perturbations must be linear. So even if
the DE contribution is non-linear, it is enough for the dark-matter
density to be dominant and its perturbation linear for the tracers
to continue to move on geodesics described by $\Phi$ and $\Psi$.
However, since there will no doubt be non-linear contributions to
the potentials, both from the DM and the DE perturbations, this should
imply that measurements of higher-order correlation functions should
be sensitive to the extra non-linearity as compared to the expected
$\Lambda$CDM result.\\

Let us pre-emptively address a few potential criticisms: our results
and approach is driven by the limitations in our ability to determine
the galaxy bias $b$ as well as the normalization $\sigma_{8}$ and
scale dependence of the dark-matter perturbations $\delta_{\text{m}}$.
There exist many, increasingly better, measurements of these quantities
e.g.~\cite{Rozo:2009jj,Tinker:2011pv,Tojeiro:2012rp,Mandelbaum:2012ay,Marin:2013bbb}.
However, they all depend in a fundamental way on assumptions necessarily
true only in $\Lambda$CDM and therefore we cannot use them.

Although non-linearities can be used to constrain the bias through
the galaxy bispectrum \cite{Marin:2013bbb}, this is a model-dependent
statement since screening mechanisms present in most dark energy models
alter the expectations. Otherwise, a measurement of bias is usually
obtained as a result of measuring $\sigma_{8}$, which is done using
either cluster counts and/or weak lensing around clusters. As we have
already pointed out, weak lensing measures the lensing potential rather
than the DM distribution, i.e.~includes a dependence on the DE perturbations.
To extract information from cluster counts, their mass is obtained
from their gas temperature---which gives information on the Newtonian
potential $\Psi$ and not on the DM mass---and spherical collapse
or N-body simulations are used, both of which depend deeply on the
model for dark energy, see e.g.~Refs~\cite{Schmidt:2009yj,Kimura:2010di,Borisov:2011fu}.
All of these issues can in principle be appropriately modeled, but
this must be done on a model-by-model basis. Without this calculation,
we cannot interpret the $\Lambda$CDM measurement of $\sigma_{8}$
as a parameter valid for other models of dark energy. \\

The tests that we have constructed in section \ref{sub:BeyQS} depend
on eliminating variables through the use of the constraints present
in the Einstein equations. Given a class of models for DE, one can
replace dynamical variables with combinations of observables. Given
the assumption that the DE is a single degree of freedom, that the
equivalence principle be satisfied and a detection of non-vanishing
anisotropic stress, we have enough information to eliminate all non-observable
quantities from the evolution equations and thus are able to form
a null test for the most general class of scalar-tensor models directly
on the observable data.

When there is no anisotropic stress then no such complete constraint
can be formed: the dependence on the unobservable $\delta_{\text{m}}$
remains. In that case, the best that can be done is to obtain a measurement
of the effective Newton's constant $Y$ on the assumption that a particular
class of models of dark energy describes the Universe and then use
a consistency relation for $Y$ to determine whether the assumption
was consistent with the data. This is a somewhat more complicated
exercise, which we have demonstrated for perfect-fluid k-\emph{essence
}models, but would also apply to imperfect-fluid models featuring
kinetic gravity braiding \cite{Deffayet:2011gz,Kimura:2010di,Pujolas:2011he}.

Tests of the type we have shown can in principle be constructed for
any other class of DE models. Indeed, we would like to argue that
one should think of the parametrized or effective approaches, such
as those of Refs~\cite{Baker:2012zs,Gubitosi:2012hu,Bloomfield:2012ff,Gleyzes:2013ooa},
as providing the dynamics of the dark energy which can be rewritten
in terms of evolution equations for the potentials. Provided they
are written in terms of parameters that on cosmological solutions
are only functions of time, a test such as those presented in section
\ref{sec:tests} would allow for putting constraints on these parameters,
or indeed would exclude setups which posses insufficient operators
to describe the data fully.

An alternative way of thinking about these tests is that, using the
measurement of $\Phi$ and $\Psi$, we are essentially reconstructing
the components of an effective combined EMT for DM and DE \cite{Ballesteros:2011cm}.
Now, given a class of models, we can extract relations between the
configuration of the degrees of freedom and the fluid variables, and
therefore between the fluid variables themselves, i.e.~anisotropic
stress, pressure, energy density. In general this is difficult, but
in some classes of models it can be done, e.g.~\cite{Battye:2012eu,Sawicki:2012re}.
However, it should be intuitively clear that if the dark energy has
more degrees of freedom, only the adiabatic modes will influence the
potentials. Internal modes, since they do not affect the gravitational
field of the DE, would only be constrainable through their impact
on the time evolution of the dark energy. In particular, one
could also form null tests on observables, but these would involve
taking additional derivatives and eliminating the internal degrees
of freedom through the appropriate equations of motion: essentially
two extra derivatives of the potentials would need to be taken for
every additional degree of freedom in the DE model class.\\

We would like to stress that our method depends on taking derivatives
of the data. Since for the purposes of this paper we are working in
the idealized case of sufficiently good data, we have not addressed
here the feasibility of this procedure in a realistic situation. We
leave this for future work.

We have also used a number of simplifying assumptions for the relation
of measurements to the potentials. For example, the Kaiser formula
\eqref{eq:Kaiser}, assumes not only linearity, but also a flat sky
and ignores near-horizon effects. As shown by e.g.~\cite{Bonvin:2011bg},
the correction involve contributions from, for example, weak lensing,
which would pollute the determination of the peculiar velocities of
galaxies from RSDs. Indeed there are also near horizon corrections
to the weak lensing, e.g.~\cite{Bernardeau:2009bm}. Since our aim
is to exploit large-scale data fully, taking into account such corrections
is a natural extension of this work. 
\begin{acknowledgments}
We would like to thank Eugeny Babichev, Nicolas Busca, Christos Charmoussis,
Valerio Marra, Federico Piazza, Glenn Starkman and Alexander Vikman
for helpful comments and suggestions. The work of L.A.~and I.S.~is
supported by the DFG through TRR33 ``The Dark Universe''. M.K.~acknowledges
funding by the Swiss National Science Foundation. M.M.~is supported
by CNPq-Brazil. I.D.S.~acknowledges STFC for financial support. 
\end{acknowledgments}
\bibliographystyle{utcaps}
\bibliography{observables}

\end{document}